\documentclass[prl,twocolumn,twoside,showpacs,superscriptaddress,floatfix]{revtex4}
\usepackage{graphics,amssymb,amsmath,epsfig}

\begin{document}
\title{Finite-size scaling in complex networks}

\author{Hyunsuk Hong}
\affiliation{Department of Physics and RINPAC, Chonbuk National University,
Jeonju 561-756, Korea}
%\affiliation{Research Institute
%of Physics and Chemistry, Jeonju 561-756, Korea}

\author{Meesoon Ha}
\affiliation{Department of Physics and RINPAC, Chonbuk National University,
Jeonju 561-756, Korea} \affiliation{School of Physics, Korea Institute for
Advanced Study, Seoul 130-722, Korea}

\author{Hyunggyu Park}
\affiliation{School of Physics, Korea Institute for Advanced Study, Seoul
130-722, Korea}

\date{\today}

\begin{abstract}
%abstract-v1
A finite-size-scaling (FSS) theory is proposed for various models in complex
networks. In particular, we focus on the FSS exponent, which plays a crucial
role in analyzing numerical data for finite-size systems. Based on the
droplet-excitation (hyperscaling) argument, we conjecture the values of the FSS
exponents for the Ising model, the susceptible-infected-susceptible model, and
the contact process, all of which are confirmed reasonably well in numerical
simulations.

\end{abstract}
%05.50.+q
%05.50.+q Lattice theory and statistics (Ising, Potts, etc.) (see also
%         64.60.Cn Order-disorder transformations and statistical mechanics of
%         model systems and 75.10.Hk Classical spin models)
%89.75.Hc Networks and genealogical trees
%05.10.-a Computational methods in statistical physics and nonlinear dynamics
%         (see also 02.70.-c in mathematical methods in physics)
%89.75.Da Systems obeying scaling laws
%64.60.Ht Dynamic critical phenomena
\maketitle

Critical phenomena in complex networks have attracted much attention recently
and traditional models in statistical physics have been examined on diverse
networks~\cite{ref:Dorogovtsev_remedy,ref:Potts_Igloi,ref:Ising_SFN_1,ref:Ising_SFN_4,ref:percolation_SFN,ref:DP_SFN,ref:synch_SFN}.
%networks~\cite{ref:Potts_Igloi,ref:Dorogovtsev_remedy,ref:Hastings,ref:Ising_SFN_1,ref:Ising_SFN_2,ref:Ising_SFN_3,ref:Ising_SFN_4,ref:XY_SW,ref:percolation_SFN,ref:DP_SFN,ref:synch_SFN}.
Interesting questions in such probes would be the existence of phase
transitions and the network-dependent critical behavior near the transition. Up
to now, various equilibrium systems such as Ising, Potts, and XY
models~\cite{ref:Dorogovtsev_remedy,ref:Potts_Igloi,ref:Ising_SFN_1,ref:Ising_SFN_4}
as well as nonequilibrium systems such as
percolation~\cite{ref:percolation_SFN}, directed percolation~\cite{ref:DP_SFN},
and synchronization models~\cite{ref:synch_SFN} have been studied by means of
the mean-field (MF) approach~\cite{ref:Potts_Igloi}, the replica
method~\cite{ref:Ising_SFN_1}, and the thermodynamic potential
hypothesis~\cite{ref:Dorogovtsev_remedy}. It is predicted that phase
transitions in complex networks exhibit mostly the {\em standard} MF critical
behavior except in highly heterogeneous scale-free (SF) networks where the
interesting heterogeneity-dependent (but still MF-type) critical behavior
appears~\cite{ref:Dorogovtsev_remedy,ref:Potts_Igloi,ref:Ising_SFN_1}.

The MF prediction has been brought up to the test by extensive numerical
simulations and passed it reasonably well in cases of less heterogeneous
networks like random, small-world, and even some SF
networks~\cite{ref:Ising_SFN_4,ref:percolation_SFN,ref:DP_SFN,ref:synch_SFN}.
However, for the highly heterogeneous networks (like the most of SF networks
found in nature), the asymptotic scaling regime could not be reached easily in
numerical tests due to huge finite size effects and consequently there is no
reasonably solid numerical analysis reported as yet. Therefore it is essential
to understand the finite-size-scaling (FSS) behavior analytically in networks,
not only to analyze numerical data for finite-size networks but also to explore
the physics of correlated size scales in networks.

In the present Letter, we propose a FSS theory for various models
in complex networks, based on a droplet-excitation (hyperscaling)
argument.  Our conjecture for the FSS exponent values is confirmed
via numerical simulations for the Ising model and the contact
process~\cite{ref:CP}.

We first start with the standard FSS theory in low dimensional systems. As a
typical example, consider the ferromagnetic Ising model. Its critical behavior
near the transition is characterized by the singular behavior in the
magnetization $m \sim \epsilon^\beta$,  the susceptibility $\chi \sim
|\epsilon|^{-\gamma}$, and the correlation length $\xi \sim |\epsilon|^{-\nu}$,
where $\epsilon$ is the reduced temperature defined by
$\epsilon\equiv(T_c-T)/T_c$.

The standard FSS theory for the singular part of the free energy $f$
reads~\cite{ref:phenomenological_FSS}
\begin{equation}
f(\epsilon,h,L^{-1}) = b^{-d}f(b^{y_{_{T}}}\epsilon, b^{y_{_{H}}}h, bL^{-1}),
\label{eq:scaling}
\end{equation}
where $b$ is the scale factor, $d$ the spatial dimension, $L$ the system linear
size, and $h$ the external field. The two scaling dimensions, $y_{_{T}}$ and
$y_{_{H}}$, determine all thermodynamic exponents such as
$\beta=(d-y_{_{H}})/y_{_{T}}$ and $\gamma=(2y_{_{H}}-d)/y_{_{T}}$. The FSS
behavior of the spontaneous magnetization is derived from Eq.(\ref{eq:scaling})
as
\begin{equation}
m=L^{-\beta y_{_{T}}}g(L^{y_{_{T}}}\epsilon), \label{eq:orderparameter_scaling}
\end{equation}
where $g$ is the scaling function and $d/y_{_{T}}=2\beta+\gamma$. If the
correlation length scales as $\xi(\epsilon,h) = b\xi(b^{y_{_{T}}}\epsilon,
b^{y_{_{H}}}h)$, the hyperscaling relation appears as
$y_{_{T}}=1/\nu$~\cite{ref:phenomenological_FSS}. In this case, the FSS
variable $L^{y_{_{T}}}\epsilon$ can be expressed as $(L/\xi)^{y_{_{T}}}$,
implying that the correlation length competes with the system size. In low
dimensions where the hyperscaling holds, the above FSS theory works perfectly
well for most equilibrium models and even for nonequilibrium ones (like
directed percolation models) with appropriate modifications~\cite{ref:Aukrust}.

In high dimensions where the MF theory becomes valid, the FSS becomes a little
bit tricky.  It has been argued that a ``dangerous irrelevant variable'' comes
into play in high dimensions which breaks the hyperscaling ($y_{_{T}}\neq
1/\nu$) and modifies the FSS~\cite{ref:dangerous_variable_1}. Nevertheless, for
block-shape samples with periodic boundary conditions, Eqs.(\ref{eq:scaling})
and (\ref{eq:orderparameter_scaling}) are still valid with the trivial MF
values of $y_{_{T}}=d/2$ and $y_{_H}=3d/4$, which correspond to the MF bulk
exponents such as $\beta=1/2$ and $\gamma=1$. The broken hyperscaling implies
that  there must be another length scale competing with the system size, rather
than the usual Gaussian correlation length.

Consider the Landau free energy of the order parameter $m$ near the transition
in the MF regime as
\begin{equation}
f(m) = - {\epsilon}m^2 + um^4 + {\cal{O}}(m^6), \label{eq:free_energy_Ising}
\end{equation}
where the Gaussian spatial fluctuation term $(\nabla m)^2$ is
ignored. In the ordered phase $(\epsilon>0)$, the free energy has
a minimum at $m=m^*\sim \epsilon^{1/2}$ with $f\sim -\epsilon^2$.
One may ask a typical size $\xi_{_T}$ of a disordered droplet
excitation out of the uniformly ordered environment. As the free
energy cost by the droplet excitation is compensated by the
thermal energy, $(\Delta f) \xi_{_T}^d\sim k_B T$, we find
$\xi_{_T}\sim \epsilon^{-\nu_{_T}}$ with
$\nu_{_T}=2/d=1/y_{_{T}}$. The Gaussian length scale diverges as
$\xi_{_G}\sim \epsilon^{-\nu_{_G}}$ with $\nu_{_G}=1/2$.

For $d>d_u=4$ (the upper critical dimension), $\xi_{_G}$ dominates over
$\xi_{_T}$, which leads to the correlation length exponent $\nu=\nu_{_G}$ and
the MF theory is valid. However, the FSS variable $L^{y_{_{T}}}\epsilon$
becomes $(L/\xi_{_T})^{y_{_{T}}}$, implying that the competing length scale is
not the dominant correlation length but the droplet size. Substituting the
linear size $L$ by the volume $N\sim L^d$,
%we have
Eq.~(\ref{eq:orderparameter_scaling}) reads
\begin{equation}
m=N^{-\beta/\bar\nu }\psi(N^{1/\bar\nu}\epsilon),
\label{eq:orderparameter_scaling4}
\end{equation}
where the FSS (droplet volume) exponent $\bar\nu=d\nu_{_T}=2$ in the MF regime.
For the general $\phi^q$ MF theory ($f=-\epsilon m^2+um^q$), we find that
$\bar\nu=d_u \nu_{_G}$ with $d_u=2q/(q-2)$, which is consistent with the
earlier result by Botet {\em et al.} for models with infinite-range
interactions~\cite{ref:Botet}.

We are now ready to explore the FSS in networks.  Networks have no space
dimensionality and may be considered as a limiting case of
$d\rightarrow\infty$.  So we expect that any model in networks displays a
MF-type critical behavior.  In particular, the MF FSS exponent $\bar\nu$ is
independent of $d$, which leads to the natural conjecture that
Eq.(\ref{eq:orderparameter_scaling4}) also applies in networks. These
predictions have been confirmed by numerical simulations for various models in
random networks, small-world networks, and complete graphs. Moreover, the
relation of $\bar\nu=d_u \nu_{_G}$ has been exploited to calculate the value of
$d_u$ via simulations in networks for complex nonequilibrium
models~\cite{ref:NohPark}.

In SF networks with the degree distribution $P(k)\sim k^{-\lambda}$, there
appears a nontrivial $\lambda$-dependent MF critical scaling for
$\lambda<\lambda_u$ (highly heterogeneous networks) while the standard MF
theory applies for $\lambda>\lambda_u$~\cite{ref:Dorogovtsev_remedy}.
Naturally, we expect a nontrivial FSS theory associated to the nontrivial MF
scaling for $\lambda<\lambda_u$. Previous studies pay attention to the MF
analysis in the thermodynamic limit and hardly discuss the FSS in the general
context.  Recently a few numerical efforts have been attempted to confirm the
MF predictions, but huge finite-size effects and the lack of the FSS theory
disallowed any decisive conclusion for highly heterogeneous
networks~\cite{ref:Ising_SFN_4,ref:synch_SFN}. Most recently, even a non-MF
scaling has been claimed for the contact
process~\cite{ref:CP_SFN_1,ref:CP_SFN_2} and a question arises whether the
cutoff in degree $k$ influences the FSS.

We start with the phenomenological MF free energy for the SF networks proposed
in~\cite{ref:Potts_Igloi,ref:Dorogovtsev_remedy}
\begin{equation}
f(m) = - {\epsilon}m^2 + um^4 + v|m|^{\lambda-1}+{\cal{O}}(m^6),
\label{eq:free_energy_Ising_SF}
\end{equation}
where the $\lambda$-dependent term originates from the singular behavior of the
higher moments of degree in SF networks. For $\lambda>\lambda_u=5$, the
$\lambda$-dependent term is irrelevant and we recover the usual $\phi^4$ MF
theory, yielding $\beta=1/2$ and $\bar\nu=2$. For $3<\lambda<5$, the
$\lambda$-dependent term becomes relevant and we find the $\phi^q$ MF theory
with $q=\lambda-1$. A simple algebra leads to $\beta=1/(\lambda-3)$ and the
free energy density in the ordered phase is $f\sim -\epsilon^{1+2\beta}$.  One
can estimate the typical droplet volume $N_{_T}\sim (\Delta f)^{-1}$, yielding
$N_{_T}\sim \epsilon^{-\bar\nu}$ with
$\bar\nu=1+2\beta=(\lambda-1)/(\lambda-3)$. By including the external field
term $hm$ in Eq.(\ref{eq:free_energy_Ising_SF}), one can show $\gamma=1$ for
all $\lambda>3$.

The results for $\beta$, $\gamma$, and the FSS exponent $\bar\nu$ for the Ising
model in SF networks are then summarized as
\begin{equation}
(\beta, \gamma, \bar\nu) = \left\{
\begin{array}{cccl}
\frac{1}{\lambda-3}, &1, &\frac{\lambda-1}{\lambda-3}& \qquad\mbox{for $3< \lambda < 5$}, \\
\frac{1}{2}, &1, &2 & \qquad\mbox{for $\lambda > 5$}.
\end{array}
\right. \label{eq:beta}
\end{equation}
For $\lambda < 3$, no phase transitions occur at finite temperatures and, at
$\lambda=5$, a multiplicative logarithmic correction is
expected~\cite{ref:Dorogovtsev_remedy}.
It is interesting to notice that a naive power counting for the $\phi^q$
local theory
with the Gaussian spatial fluctuation term $(\nabla m)^2$ yields the same
result for $\bar\nu$ by using the relation
of $\bar\nu=d_u \nu_{_G}$~\cite{ref:SIS+cutoff}.  Our
conjecture for $\bar\nu$ bears no reference to the degree cutoff $k_c$ caused
by the finite system size $N$.  We will argue later that the cutoff is
irrelevant if it is not too strong:
$k_{c} > N^{1/\lambda}$~\cite{ref:CP_SFN_2}.

We check our conjecture via numerical simulations. Two typical SF networks are
considered, namely the static model~\cite{ref:SNUstatic_SFN} and the
uncorrelated configuration model (UCM)~\cite{ref:UCM_SFN}.  As these networks
have different degree cutoffs (natural cutoff $k_c\sim N^{1/(\lambda-1)}$
versus forced sharp cutoff $k_c\sim N^{1/2}$) in finite systems, one may look
for a possibility of the cutoff-dependent FSS behavior if any.  It turns out
that both cutoffs are not strong enough to influence the FSS for $\lambda>2$.

\begin{figure}
\includegraphics[width=0.45\textwidth]{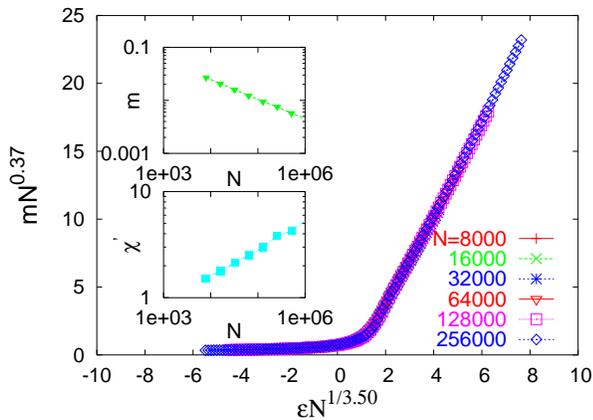}
\caption{ \label{fig:Ising-m} (Color online) Data collapse of $m$ for the
static network with $\lambda=3.87$, using $\beta/\bar{\nu}=0.37$ and
$\bar{\nu}=3.5$. Insets: double logarithmic plots of the critical decay of
%$m$ (upper) and $\chi^\prime$ (lower) against system size with slopes
%$\beta/\bar{\nu}=0.37(5)$ and $\gamma^\prime/\bar{\nu}=0.26(4)$, respectively.}
$m$ and $\chi^\prime$ against $N$ with slopes $\beta/\bar{\nu}=0.37(5)$ and
$\gamma^\prime/\bar{\nu}=0.26(4)$.}
\end{figure}

We performed Monte Carlo simulations at various values of $\lambda$ up to
$N=10^7$.  We measure the magnetization $m$,  the fluctuation $\chi^\prime=
N(\Delta m)^2$, and the Binder cumulant $B$ and average over
$\sim 10^3$ network realizations. The transition temperature $T_c$ is
estimated by the asymptotic limit of the crossing points of $B$ for
successive system sizes as well as of the peak points of $\chi^\prime$.
At criticality, Eq.(\ref{eq:orderparameter_scaling4}) leads
to $m\sim N^{-\beta/\bar\nu}$ and
similarly $\chi^\prime\sim N^{\gamma^\prime/\bar\nu}$ with $\chi^\prime\sim
|\epsilon|^{-\gamma^\prime}$ in the thermodynamic limit.  This power-law
behavior in $N$ provides an alternative check for the criticality as well as
the estimates for the exponent ratios.  In equilibrium systems, the
fluctuation-dissipation theorem guarantees $\gamma^\prime=\gamma$. By
collapsing the data over the range of temperatures, we  estimate the value of
the FSS exponent $\bar\nu$. Our numerical data for $m$ and $\chi^\prime$
collapse very well for all values of $\lambda$ in both static and UCM networks.
In Fig.~\ref{fig:Ising-m}, the data collapse is shown for $\lambda=3.87$ in
static networks.  We summarize in Table I the numerical estimates for
$\beta/\bar\nu$, $\bar\nu$, and $\gamma^\prime/\bar\nu$ at various values of
$\lambda$ in static and UCM networks. All data agree reasonably well with our
predictions.

%%%%%%%%%%%%%%%%%%%%%%%%%%%%%%%%%%%%%%%%%%%%%%%%%%%%%%%%%%%%%
%%%%%%%%%%%%%%%%%%%%%%%%%%%%%%%%%%%%%%%%%%%%%%%%%%%%%%%%%%%%%

We also measure the degree-dependent quantities like the magnetization on
vertices of degree $k$, $m_k$, and its fluctuation $(\Delta m_k)^2$.  These
quantities are found to satisfy a scaling relation with the scale variable
$kN^{-1/\lambda}$ (not shown here)~\cite{ref:SIS+cutoff}.
For $k>N^{1/\lambda}$, the distribution
$P(k)$ becomes almost flat for each realization of networks and the degree
exponent $\lambda$ loses its identity.  Therefore, vertices of such a high
degree contribute in a trivial way and the cutoff beyond this range
($k_c>N^{1/\lambda}$) should not be distinguishable~\cite{ref:CP_SFN_2}.
This argument is
supported by our numerical results which cannot differentiate the FSS scaling
in the static and UCM networks.
%%%%%%%%%%%%%%%%%%%%%%%%%%%%%%%%%%%%%%%%%%%%%%%%%%%%%%%%%%%%%%%%%%%
\begin{table}
\caption{Critical exponents of the  Ising model on the static and the UCM
networks, and the CP on the UCM networks, compared with our MF predictions.}
\begin{tabular}{cc|ccc}
\hline\hline
& & &Ising&\\
network&$\lambda$&$\beta/\bar{\nu}$&$\bar{\nu}$&$\gamma^\prime/\bar{\nu}$\\
\hline
MF&$\lambda>5$& 1/4&2&1/2\\
  &$3<\lambda<5$& $\frac{1}{\lambda-1}$&$\frac{\lambda-1}{\lambda-3}$&$\frac{\lambda-3}{\lambda-1}$\\
\hline
static&7.08&0.26(4)&2.0(2)&0.45(5)\\
      &4.45&0.28(2)&2.4(2)&0.45(3)\\
      &3.87&0.37(5)&3.5(3)&0.26(4)\\
\hline
UCM&6.50&0.24(4)&2.0(2)&0.51(5)\\
   &4.25&0.31(1)&2.5(1)&0.39(1)\\
   &3.75&0.38(6)&3.9(2)&0.24(3)\\
\hline\hline
%\end{tabular}
%\begin{tabular}{cc|cccccc}
& & &CP&\\
network&$\lambda$&$\beta/\bar{\nu}$&$\bar{\nu}$&$\gamma'/\bar{\nu}$\\
\hline
MF&$\lambda>3$& 1/2&2&0\\
  &$2<\lambda<3$& $\frac{1}{\lambda-1}$&$\frac{\lambda-1}{\lambda-2}$&$\frac{\lambda-3}{\lambda-1}$\\
\hline
UCM &4.0&0.49(1)&2.1(1)&0.00(5)\\
    &2.75&0.58(1)&2.4(1)&-0.16(2)\\
    &2.25&0.78(1)&4.0(5)&-0.55(5)\\
%\hline
%& & &CP $(h\ne0)$&\\
%network&$\lambda$&$\beta/\sigma$&$\sigma/\bar{\nu}$&$\gamma'/\sigma$\\
%\hline
%MF&$\lambda>3$& 1/2&1&0\\
%  &$2<\lambda<3$& $\frac{1}{\lambda-1}$&1&$\frac{\lambda-3}{\lambda-1}$\\
%\hline
%UCM &4.0&0.51(2)&1.0(1)&-0.04(4)\\
%    &2.75&0.59(2)&1.0(2)&-0.18(4)\\
%    &2.25*&0.70(4)&1.3(1)&-0.33(8)\\
%&&&\\
\hline\hline
\end{tabular}
\end{table}
%%%%%%%%%%%%%%%%%%%%%%%%%%%%%%%%%%%%%%%%%%%%%%%%%%%%%%%%%%%%%%%%%%%

Now we move to a typical model exhibiting a nonequilibrium phase transition,
namely the directed percolation (DP) system~\cite{ref:Nonequil_Hinrichsen}. It
has been well known that most of nonequilibrium models showing an
absorbing-type phase transition belong to the DP universality class. Among such
models, we here consider the contact process (CP) and the
susceptible-infected-susceptible (SIS) model~\cite{ref:DP_SFN}.

The CP is an interacting particle model on a lattice.  A particle creates
another particle in one of its neighboring sites with rate $p$ and a particle
annihilates with rate 1. In the SIS model, the particle creation  is attempted
in all neighboring sites. A particle-particle interaction comes in through
disallowance of multiple occupancy at a site.  As $p$ increases, the system
undergoes a phase transition at $p_c$ from a quiescent vacuum (absorbing) phase
to a noisy many-particle (active) phase in the steady state. Near the absorbing
phase transition, the order parameter (particle density) $\rho\sim
\epsilon^\beta$, the fluctuations $\chi^\prime=N(\Delta\rho)^2\sim
\epsilon^{-\gamma^\prime}$, the susceptibility $\chi\sim |\epsilon|^{-\gamma}$,
the correlation length $\xi\sim |\epsilon|^{-\nu}$, the relaxation time
$\tau\sim|\epsilon|^{-\nu_t}$, and the survival probability $P_s\sim
\epsilon^{\beta^\prime}$ with the reduced coupling constant
$\epsilon=(p-p_c)/p_c$. It is known that $\beta=\beta^\prime$ due to the
time-reversal symmetry in the DP systems~\cite{ref:Nonequil_Hinrichsen} and
$\gamma^\prime\neq \gamma$ in general nonequilibrium systems.

Consider the droplet (cluster) excitation starting from a localized seed in the
absorbing phase. The average space-time size $S$ of a cluster is estimated as
\begin{equation}
S\sim \tau_\ell ~ \xi_c^d \sim |\epsilon|^{-\sigma},
\label{eq:clustersize}
\end{equation}
where $\tau_\ell$ and $\xi_c$ are the average lifetime and typical size of a
droplet, respectively. Usually $\tau_\ell$ diverges near the transition as
$\tau_\ell\sim |\epsilon|^{-\nu_t+\beta^\prime}$ for
$\nu_t>\beta^\prime$~\cite{ref:Nonequil_Hinrichsen}, but $\tau_\ell$ is a
${\cal O}(1)$ constant otherwise.  In the MF regime, it is shown later that the
latter always applies. The droplet size diverges as $\xi_c \sim
|\epsilon|^{-\nu_{_T}}$,  which leads to $\sigma=d\nu_{_T}+\mbox{ max}
\{\nu_t-\beta^\prime, 0\}$. It is well known that the susceptibility is
proportional to the cluster mass, which yields
$\gamma=\sigma-\beta$~\cite{ref:Nonequil_Hinrichsen}. Finally, we arrive at the
generalized exponent relation as
\begin{equation}
\gamma=d\nu_{_T}-\beta +\mbox{ max} \{\nu_t-\beta^\prime, 0\}.
\label{eq:GFD}
\end{equation}
The fluctuation exponent $\gamma^\prime$ satisfies the standard
hyperscaling relation as $\gamma^\prime=d\nu_{_T}-2\beta$.

In SF networks, we propose a phenomenological modification of the
MF Langevin equation describing the DP models, similar to the free
energy modification of the Ising model in
Eq.(\ref{eq:free_energy_Ising_SF}):
\begin{equation}
\frac{d}{dt}\rho(t) = \epsilon \rho - b\rho^2 - d\rho^{\theta-1} +
\sqrt{\rho}\eta(t), \label{eq:Dorogovtsev-DP}
\end{equation}
where $\rho(t)$ is the particle density at time $t$ and $\eta(t)$ is a Gaussian
noise.  Our modification to the standard MF theory comes in by the third
$\rho^{\theta-1}$ term and it is straightforward to show that the exponent
$\theta=\lambda$ for the CP and $\theta=\lambda-1$ for the SIS.

By dropping the noise term, one may easily get the MF steady-state solution for
$\rho$. We find that $\beta=1$ for $\theta>3$ and $\beta=1/(\theta-2)$ for
$2<\theta<3$. For $\theta< 2$, there is no phase transition at finite $p$. The
same result may be obtained from the well-established the $k$-dependent
noiseless MF theory~\cite{ref:DP_SFN}.  Moreover, the time-dependent solution
leads to $\nu_t=1$ and the inclusion of a weak external field leads to
$\gamma=1$ for all $\theta>2$. Notice that $\nu_t\le \beta=\beta^\prime$.

Utilizing the exponent relation of Eq.(\ref{eq:GFD}), we summarize the results
for $\beta$, $\gamma$, and the FSS exponent $\bar\nu=d\nu_{_T}=1+\beta$  as
\begin{equation}
(\beta, \gamma, \bar\nu) = \left\{
\begin{array}{cccl}
\frac{1}{\theta-2}, &1, &\frac{\theta-1}{\theta-2}& \qquad\mbox{for $2< \theta < 3$}, \\
1, &1, &2 & \qquad\mbox{for $\theta > 3$},
\end{array}
\right. \label{eq:beta-DP}
\end{equation}
where $\theta=\lambda$ for the CP and $\theta=\lambda-1$ for the
SIS. For both cases,  $\nu_t=1$, $\sigma=\bar\nu$, and
$\gamma^\prime=1-\beta$. As in the Ising model, a naive power
counting for the local Langevin equation with the spatial
fluctuation term $\nabla^2 \rho$ yields the same result for
$\bar\nu$ by using $\bar\nu=d_u \nu_{_G}$~\cite{ref:SIS+cutoff}.

%%%%%%%%%%%%
%%%%%%%%%%%%%%%%%%%%%%%%%%%%%%%%%%%%%%%%%%%%%%%%%%%%%%%%%%%%%%%%%%%%%%%%
\begin{figure}[t]
\includegraphics[width=0.45\textwidth]{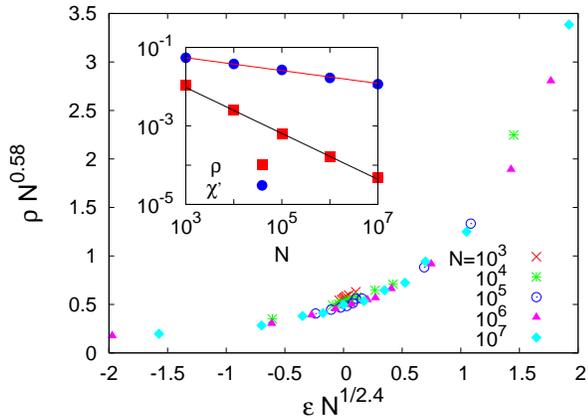}
\caption{ \label{fig:cp-m} (Color online) Data collapse of  $\rho$ for the UCM
network with $\lambda=2.75$, using $\beta/\bar{\nu}=0.58$ and $\bar{\nu}=2.4$.
Inset: double logarithmic plots of the critical decay of $\rho$  and $\chi'$
against $N$ with slopes $\beta/\bar{\nu}=0.58(1)$ and
$\gamma^{\prime}/\bar{\nu}=-0.16(2)$.}
\end{figure}
%%%%%%%%%%%%%%%%%%%%%%%%%%%%%%%%%%%%%%%%%%%%%%%%%%%%%%%%%%%%%%%%%%%%%%%%
We performed numerical simulations for the CP on the UCM networks at various
values of $\lambda$ up to $N=10^7$. We measure  $\rho$ and  $\chi^\prime=
N(\Delta \rho)^2$. The transition point $p_c$ is estimated by the power-law
temporal dependence of $\rho$ and also by the power-law size dependency of
$\rho$ in the steady state. At criticality, the temporal dependence is given as
$\rho\sim t^{-\beta/\nu_t}$ and, in the steady state, $\rho\sim
N^{-\beta/\bar\nu}$ and $\chi^\prime\sim N^{\gamma^\prime/\bar\nu}$. By
collapsing the off critical data in the steady state, we can estimate the value
of $\bar\nu$.

In Fig.~\ref{fig:cp-m}, the data collapse is shown for $\lambda=2.75$ in the
UCM networks. Our estimates for the exponents summarized in Table I agree well
with our predictions, in general. Our data become a little bit weaker close to
$\lambda=2$ where high heterogeneity in networks yields big corrections to
scaling. We also confirmed the relation of $\sigma=\bar\nu$ by investigating
the external field effect at criticality. Simulations were also performed on
the static networks, where no cutoff dependence is found. The SIS model shows
bigger finite-size corrections, which hinder us to estimate the transition
point accurately. Nevertheless, our estimates for the steady-state exponent
ratios are consistent with our predictions within errors~\cite{ref:SIS+cutoff}.

%%%%%%%%%%%%%% Summary %%%%%%%%%%%%%%%%
In summary, we have explored the finite-size scaling (FSS) behavior in complex
networks. Based on the droplet-excitation argument, we conjectured the FSS
exponent values, which have been confirmed by extensive numerical simulations
on the static and UCM networks. Real networks found in many biological,
economical, and social systems are highly heterogeneous and also quite small in
size. Our results will provide the essential information on analyzing the data
on these networks.

This work was supported by Korea Research Foundation Grant funded by
the Korean Government (MOEHRD) (KRF-2006-331-C00123) and by the BK21
project.

\end{document}